\documentclass[11pt]{article}
\usepackage{geometry}
\usepackage{cite}
\usepackage{comment}
\usepackage{amsmath}
\usepackage{amssymb}
\usepackage{amsfonts}
\usepackage{graphicx}
\usepackage{caption}
\usepackage{subcaption}
\usepackage{float}
\usepackage[usenames,dvipsnames]{xcolor}
\usepackage{tikz}
\usepackage{pstricks}
\usepackage{lipsum}
\usepackage{appendix}
\usepackage{authblk}
\usepackage{verbatim}
\usepackage{setspace}
\usepackage{gensymb}
\renewenvironment{thebibliography}[1]{
  \begin{oldthebibliography}{#1}
    \setlength{\itemsep}{0em}
    \setlength{\parskip}{0em}
}
{
  \end{oldthebibliography}
}
\begin{document}
\date{}
\author{Sandip Chowdhury \thanks{sandipc@iitk.ac.in}}
\author{Kunal Pal \thanks{kunalpal@iitk.ac.in}}
\author{Kuntal Pal \thanks{kuntal@iitk.ac.in}}
\author{Tapobrata Sarkar \thanks{tapo@iitk.ac.in}}
\affil{Department of Physics, \\ Indian Institute of Technology, \\ Kanpur 208016, India}
\title{Quantum potential in bouncing dust collapse with a negative cosmological constant}
%%%%%%%%%%%%%%%%%%%%%%%%%%%%%%%%
\maketitle
\begin{abstract}
In the functional Schrodinger formalism, we obtain the wave function describing collapsing dust in an anti-de Sitter
background, as seen by a co-moving observer, by mapping the resulting variable mass Schrodinger equation to that 
of the quantum isotonic oscillator. 
Using this wave function, we perform a causal de Broglie-Bohm analysis, and obtain the corresponding quantum potential. 
We construct a bouncing geometry via a disformal transformation, incorporating quantum effects. 
We derive the external solution that matches with this smoothly, and is also quantum corrected.  
Due to a pressure term originating from the  quantum potential, an initially collapsing solution with a negative 
cosmological constant bounces back after reaching a minimum radius, and thereby avoids the classical singularity 
predicted by general relativity.

\end{abstract}
\maketitle

\section{Introduction}	

One of the outstanding issues that spur interest in general relativity (GR) is the nature of quantum effects in 
the regime of strong gravity. Collapsing scenarios in GR provide a unique laboratory for such studies. Indeed, 
gravitational collapse has been extremely well studied in GR (see, for example the reviews \cite{JoM}, \cite{Malafarina} 
and references therein), as well as in alternative gravity scenarios \cite{Cembranos}, \cite{Goswami}, \cite{Tapo1}, 
and it is well known that the end stage of a generic collapsing star can either be a black hole or a naked singularity. 
The pioneering work of Penrose and Hawking \cite{HP} shows that in such a collapsing scenario, the 
occurrence of singularities is inevitable in classical GR. However, an important question is whether this signals a lack of our
understanding of the ``correct'' theory in the strong gravity regime. Since a singularity essentially indicates a 
breakdown of the underlying theory itself, the widespread belief is that a consistent quantum theory of 
gravity should be non-singular in strong gravity regimes. 

In GR, the metric plays the role of a dynamical field, and hence in a quantum theory of gravity, this is to be treated as 
a quantum operator, with possible quantization conditions on space and time as well. This of course poses a formidable
challenge, although seminal works have appeared on related topics over the last few decades (for a sampling of the 
literature, see the textbooks \cite{CarlipQG}, \cite{RovelliQG}, \cite{KieferQG}, \cite{ThiemannQG}, \cite{BojowaldQG} and references therein).
In view of this, a major line of research has been to carry out semi-classical analyses (where the metric is treated at 
the classical level), with the hope that the results obtained from these will be generic. In particular, as explained 
in \cite{KieferQG}, in this line of approach, one often uses the functional
Schrodinger equation that arises from the minisuperspace version of  the Wheeler-De Witt (WDW) \cite{dw} equation, 
in a first quantized approach. This has been used efficiently to analyze different aspects of quantum 
mechanical effects on gravitational collapse and Hawking radiation.

In a variety of cases, such analyses enable us to conclude that the classical singularity is removed. This is based on the
De Witt criterion, which states that a sufficient (but not necessary) criterion for avoiding the classical  singularity is that 
the wave function should vanish at the singularity (see, e.g \cite{CK}). Our purpose in this paper, on the other hand, 
will be to compute a quantum corrected metric\footnote{See 
 the recent works \cite{New1}, \cite{New2}, \cite{New3} where the authors constructed a quantum version of the  
 Oppenheimer-Snyder dust collapse model.}. A well known method to do this is to 
solve the quantum corrected Einstein equations, taking into account of the backreaction of the quantum fields \cite{Birrell}. 
Here, we will explore another useful way to take into account quantum effects, by using an alternative interpretation of 
quantum mechanics, namely the causal de Broglie-Bohm (dBB) theory \cite{PH}. In the dBB formalism, quantum particles move 
in definite trajectories acted upon by a quantum potential, in addition to an external classical potential. 

With this version of quantum mechanics, one can take into account the quantum effects by doing a 
conformal \cite{RC} or disformal {\cite{QM}}  transformation of the original singular classical metric. 
One can then argue that the new metric contains quantum effects in the conformal or disformal degrees of freedom,
and that the quantum metric is singularity free. Here, we give such a dBB interpretation of the functional  
Schrodinger  equation, which is equivalent to two equations, namely, a continuity equation for the probability density, 
and a modified quantum Hamilton-Jacobi (HJ) equation. By comparing with the classical HJ equation, and using a particular
linear superposition of the exact wavefunctions obtained for dust collapse with a negative cosmological constant 
as the initial state of the system, we obtain the expression for the quantum potential. We show that the effect of the 
quantum potential is maximum close to the classical singularity and due to its effect, the collapsing shell is never able 
to reach the singularity. Instead, it bounces back after a certain minimum radius.
While it is well known that in an usual GR scenario of dust  collapse with a negative cosmological constant, the final
outcome always results in the classical singularity when the scale factor inevitably vanishes, here we show that the quantum effects 
create a pressure (the energy momentum tensor including the contribution form the quantum potential behaves like 
a perfect fluid) so that  the dust ball can avoid the singularity. The quantum metric with such a nonsingular scale factor 
is obtained through a disformal transformation. Here, as an approximation, we have ignored radiation effects. 
In general, there might be two issues here. Firstly, the collapsing matter can itself radiate, and secondly, there will be 
Hawking radiation. Both these effects are ignored in this work. 	

This paper is organized as follows. In the next section \ref{secFRW}, we consider dust collapse in the presence of 
a negative cosmological constant as viewed by a co-moving observer, 
in the functional Schrodinger formalism. We show how this leads to a variable mass Schrodinger equation. 
The latter problem is then cast in the form of the quantum isotonic oscillator, and we find explicit solutions for the wave function. 
In section \ref{secdBB}, we first undertake a dBB analysis of the Schrodinger equation, and obtain the quantum
potential. We then construct a modified quantum version of the metric that gives rise to a Hamiltonian that 
contains the quantum potential as a correction. The time evolution of this metric is then studied to demonstrate the
effect of the quantum potential in the collapse process. The horizon structure of our solution is
discussed in section \ref{Hor}. We conclude with a discussion of our results in section \ref{Conc}. 	
%This paper contains an appendix that lists some of the important details of the quantization procedure with 
%a variable mass. 

\section{Functional Schrodinger formalism and dust collapse}
\label{secFRW}

We consider a spherically symmetric collapsing dust, where the interior metric is smoothly matched through a timelike hypersurface
with an external Schwarzschild de Sitter (SdS) or Schwarzschild anti de Sitter (SAdS)  solution, depending on the sign of the 
cosmological constant. 
The standard junction conditions of 
general relativity \cite{EP} are assumed to hold. 
Let us suppose that our collapsing dust is represented by the Hamiltonian  $\mathcal{H}$, and the total 
wave functional of the system is by $\Psi[X^{\alpha},g_{\mu\nu},O]$, 
where $X^{\alpha}$ and $O$ denotes the degrees of freedom  
of the collapsing system, and the observer, respectively \cite{VSK}. Then the celebrated WDW equation \cite{dw} 
is given by ${\mathcal H}\Psi=0$.
If we write the total Hamiltonian in two parts, which corresponding of the system and observer, i.e.,
${\mathcal H}={\mathcal H}_{sys} + {\mathcal H}_{obs}$,
and assume that the interaction between the system and observer is weak (i.e., the observer does not affect the 
system significantly), then the WDW equation implies the 
Schrodinger equation that corresponds to the evolution of the system ($\hbar = \frac{h}{2\pi}$, where $h$
is Planck's constant)
with respect to the observer's time $t$  \cite{VSK}, \cite{GS},
\begin{equation}\label{sch1}
{\mathcal H}_{sys}\Psi_{sys}=i\hbar\frac{\partial\Psi_{sys}}{\partial t}~.
\end{equation}
Given ${\mathcal H}_{sys}$, the  solutions of this equation are interpreted as the wave functions of the system $\Psi_{sys}$. 
From now on, in order not to clutter
the notation, we will drop the subscript, and use the notations ${\mathcal H}$ and $\Psi$ for the Hamiltonian and the wave
function of the system. 

For a generic spherically symmetric collapsing interior solution, we have, in comoving coordinates  
\begin{equation}
ds^2 = -e^{2\nu(r,t)}c^2dt^2 + e^{2\psi(r,t)}dr^2 + Q(r,t)^2d\Omega^2~.
\label{gencoll}
\end{equation}
Here, $Q(r,t)$ denotes the comoving radius of the collapsing sphere, $\nu(r,t)$ and $\psi(r,t)$ are two arbitrary
real functions of $r$ and $t$, $c$ is the speed of light, 
and $d\Omega^2 = d\theta^2 + \sin^2\theta d\phi^2$. The Misner-Sharp mass 
function \cite{MisnerSharp1}\cite{MisnerSharp2}, is given by 
\begin{equation}
M(r,t) = \frac{Q(r,t)c^2}{G}\left[1-e^{-2\psi(r,t)}Q'(r,t)^2 + e^{-2\nu(r,t)}\frac{{\dot Q}(r,t)^2}{c^2}\right]~,
\label{MisnerSharp}
\end{equation}
with $G$ denoting the gravitational constant and a prime and an overdot indicating a derivative with respect to 
the spatial and the temporal coordinates, respectively. The $tt$ and $rr$ components of the Einstein's equations with a cosmological 
constant $\Lambda$ for a perfect fluid source can be written in terms of the Misner Sharp mass function as
\begin{equation}
\frac{M'}{Q^2Q'} = \rho_{mat}  + \frac{\Lambda c^2}{G}~,~~ \frac{{\dot M}}{Q^2{\dot Q}} =  
-\frac{p_{mat}}{c^2} + \frac{\Lambda c^2}{G}~,
\label{Eingen}
\end{equation}
where the subscript ``$mat$'' indicates the matter part of the contribution to the density $\rho$ and the
pressure $p$, with the perfect fluid  energy-momentum tensor having components 
$T^{\mu}_{\nu} = {\rm diag}~(-\rho c^2, p, p, p)$. While writing the Einstein's equations, we will neglect an
unimportant factor of $8\pi$ throughout. 

We will be interested in the Friedman-Robertson-Walker model, for which the metric inside the collapsing dust sphere is given by
\begin{equation}
ds^{2}_{-}=-c^2d\tau^{2}+a^{2}(\tau)\Big[\frac{dr^{2}}{1-kr^{2}}+r^{2}d\Omega^{2}\Big]~.
\label{FRW}
\end{equation}
Here, $a(\tau)$ is the scale factor, and $\tau$ is the comoving time.  
Since the pressure of matter is zero, this is also the proper time measured by such an observer. Here, $k$ is a constant,
which can take values $(-1,0,1)$. The metric outside the collapsing sphere 
is given by the SdS or SAdS metric,
\begin{equation}
ds^{2}_{+}=-f dt^{2}+f^{-1} d\tilde{r}^{2}+\tilde{r}^{2}d\Omega^{2}~,~~ f
=1-\frac{2G\mathcal{M}}{c^2\tilde{r}}-\frac{\Lambda}{3}\tilde{r}^{2}~.
\end{equation}
Where $\mathcal{M}$ is the Schwarzschild mass.
The interior metric is matched with this solution through a timelike hypersurface at $r=constant$. In this case, the mass function of eq.(\ref{MisnerSharp}) is given by 
\begin{equation}
M(r,a(\tau))=\frac{ar^3c^2}{G}\left(\frac{{\dot a}^2}{c^2} + k\right)~.
\label{Junc2}
\end{equation}
Then the Einstein's equations of eq.(\ref{Eingen}) are straightforwardly given as 
\begin{equation}
\frac{M'}{a^3r^2}=\rho_{mat} + \frac{\Lambda c^2}{G}~,~~ \frac{{\dot M}}{r^3a^2{\dot a}} = \frac{\Lambda c^2}{G}~,
\label{Eins}
\end{equation}
here an overdot indicates derivative with respect to proper time $\tau$. 
The second equation of eq.(\ref{Eins}) can be immediately integrated to give 
$M(r,\tau)=\Lambda c^2 a^3 r^3/(3 G) + r^{3} \mathcal{H}$, where  $\mathcal{H}$ is a constant and we have fixed the radial 
dependence of this term by demanding that the energy density should be regular at the center of the dust cloud $r=0$ at the 
start of the collapse. Then, upon using eq.(\ref{Junc2}), we obtain a conserved quantity 
\begin{equation}
\frac{ac^4}{G}\left(\frac{{\dot a}^2}{c^2} + k\right) - \frac{\Lambda c^4 a^3}{3 G} \equiv {\mathcal H} ~.
\label{conserved}
\end{equation}
From the point of view of the comoving observer, ${\mathcal H}$ is a constant of motion. Now,  
following \cite{VSK}, \cite{GS} (see also \cite{vs}, \cite{ss}), we identify ${\mathcal H}$ as the Hamiltonian of the system,
in the comoving frame. The effective action and the generalized momenta $p_{a}$ are then given by
\begin{equation}
\mathcal{S}_{eff}=\int d\tau \Big[\frac{ac^4}{G}\left(\frac{{\dot a}^{2}}{c^2}-k\right) 
+ \frac{\Lambda c^4 a^3}{3G} \Big]~, ~~ p_{a}=\frac{2c^2a{\dot a}}{G}~,
\label{Se1}
\end{equation}
in terms of which we can write the Hamiltonian as 
\begin{equation}
\mathcal{H}\left(a,p_{a}\right)=\frac{p_{a}^{2}}{2(2ac^2/G)}+\left(\frac{kac^4}{G} - \frac{\Lambda c^4 a^3}{3G}\right)~.
\label{H21}
\end{equation}

Our task then boils down to solving the functional Schrodinger equation of eq.(\ref{sch1}) for this Hamiltonian.
Note that this can be viewed as the Hamiltonian of a particle with a position dependent mass $m(a)=2ac^2/G$ moving in a 
potential $V(a)=kac^4/G - \Lambda c^4 a^3/(3G) $. 
In the quantum version of the theory, since position and momentum does not commute, 
we need to be careful in writing the corresponding Schrodinger equation for this Hamiltonian. 
Here the mass $m(a)$ becomes a position dependent operator, $\textbf{M(\textbf{X})}=2\textbf{X}$, and hence does not 
commute with the momentum operator\footnote{$\textbf{X}$ and $ \textbf{P}$ denotes the operator form 
of the position $a$ and momentum $p_{a}$ respectively. Since we are in position space $\textbf{X}$ is just $a$ and 
$\textbf{P}=-i\hbar\frac{\partial}{\partial a}$. We will use boldfaced letters here to denote the quantum operators.} $\textbf{P}$. 
If we simply replace $m$ with $\textbf{M}(\textbf{X})$ in the operator form of the Hamiltonian eq.(\ref{H21}), then it will become 
non-Hermitian. Thus we need to do a careful ordering of $\textbf{P}$ and $\textbf{X}$.
For the moment, will work in units $G=c=1$. These will be put back as and when appropriate, later. 

One way to obtain a proper ordering of the momentum and the position dependent mass, such that the kinetic part of the 
resulting Hamiltonian is Hermitian, is to introduce the following 
two parameter family of Hamiltonians with position $\textbf{X}$, momentum $\textbf{P}$ and position
dependent mass $\textbf{M(\textbf{X})}$, with the kinetic part given by \cite{JL}
\begin{equation}
\mathcal{H}_{kin}~~~=\frac{1}{4}(\textbf{M}^{\alpha}\textbf{P}\textbf{M}^{\beta}\textbf{P}\textbf{M}^{\gamma}+\textbf{M}^{\gamma}
\textbf{P}\textbf{M}^{\beta}\textbf{P}\textbf{M}^{\alpha})
= \frac{1}{2}\textbf{P}\frac{1}{\textbf{M(\textbf{X})}}\textbf{P}+\frac{1}{2}(\alpha+\gamma+\alpha\gamma)
\frac{\textbf{M}^{\prime 2}}{\textbf{M}^{3}}-\frac{1}{4}(\alpha+\gamma)\frac{\textbf{M}^{\prime\prime}}{\textbf{M}^{2}}~.
\label{HK1}
\end{equation}
Here,  we need to impose the constraint $\alpha+\beta+\gamma=-1$. 
It has been shown in \cite{JL}, by using the method of instantaneous Galilean transformations,  
eq.(\ref{HK1}) is  the most general class of kinetic Hamiltonians.
If we choose $\alpha=\gamma=0, \beta=-1$, then  the kinetic part of the Hamiltonian reduces to the simple form
\begin{equation}\label{hkin}
\mathcal{H}_{kin}=\frac{1}{2}\textbf{P}\frac{1}{\textbf{M(\textbf{X})}}\textbf{P}.
\end{equation}  
Note that this choice is not unique, but the most suitable for our purposes. 
Now using the operator form of the momentum, the Schrodinger equation is
\begin{equation}
-\frac{\hbar^2}{2}\frac{\partial}{\partial a}\Big[\frac{1}{m(a)}\frac{\partial \Psi   (a,\tau)}{\partial a}\Big]+ V(a)\Psi(a,\tau)
= i\hbar\frac{\partial \Psi(a,\tau)}{\partial \tau}~.
\label{eq1}
\end{equation}
Since the potential $V(a)$ or the  mass $m(a)$ do not explicitly depend on 
time, we can separate  the time independent part of eq.(\ref{eq1}), which, for energy $E$ is given by
\begin{equation}\label{sch3}
-\frac{\hbar^2}{2}\frac{d}{d a}\Big[\frac{1}{m(a)}\frac{d\psi(a)}{d a}\Big]+ V(a)\psi(a) = E\psi(a), ~~~ \Psi(a,\tau)=\psi(a)e^{-\frac{iE\tau}{\hbar}}.
\end{equation}
To solve the time-independent Schrodinger equation, we introduce the following coordinate transformations in 
terms of a real coordinate $z$ (see e.g. \cite{BG})
\begin{equation}
a=f(z),~ \psi(a)=g(z)\phi(z)~,~~ m(a)=m[f(z)]=\tilde{m}(z)~.
\label{cond1}
\end{equation}
If we substitute these transformations in eq.(\ref{sch3}), we can make the resulting equation a constant mass (say $m_{0}$) 
Schrodinger equation (with a potential different  from $V(a)$) by making the following choices
\begin{equation}
f^{\prime}(z)=\sqrt{\frac{m_{0}}{\tilde{m}(z)}}~,~~ g(z)={\mathcal C}\sqrt{f^{\prime}(z)\tilde{m}(z)} = 
{\mathcal C}(m_{0}\tilde{m})^{(1/4)}~.
\label{tr1}
\end{equation}
Here, ${\mathcal C}$ is an integration constant. If the original wave function $\psi(a)$ is square integrable, 
then it can be shown that  the new wave function $\phi(z)$ will also be square integrable if we take ${\mathcal C}=1/\sqrt{m_{0}}$.
The first relation of eq.(\ref{tr1}) can be inverted to get the relation between the coordinates and hence the mapping function  
\begin{equation}\label{atoz}
z=\frac{1}{\sqrt{m_{0}}}\int_{0}^{a}\sqrt{m(a)} da =f^{-1}(a). 
\end{equation}
With eqs.(\ref{cond1}) and (\ref{tr1}), eq.(\ref{sch3}) reduces to
\begin{equation}\label{cmSc}
-\frac{\hbar^2}{2m_{0}}\frac{d^{2}\phi}{dz^{2}}+\bigg(V_{m}(a)+V(a)\bigg)_{a(z)}\phi(z)=E\phi(z)~,
\end{equation}
where $V_{m}$ is the mass dependent part of the potential, given by
\begin{equation}\label{Vm}
V_{m}=\frac{\hbar^2}{32}\Bigg[\frac{1}{m(a)}\Bigg(7\left(\frac{m^{\prime}(a)}{m(a)}\right)^{2}-
4\left(\frac{m^{\prime\prime}(a)}{m(a)}\right)\Bigg)\Bigg]_{a(z)}.
\end{equation}
Since this potential arises out of the first term in eq.(\ref{sch3}), it is proportional to $\hbar^{2}$, 
and can be named as the quantum part of the total potential.
Now, eq.(\ref{cmSc}) describes a particle of constant mass $m_{0}$ moving in a total potential 
$\Omega(z)=(V_{m}+V)_{a(z)}$. If we can find the solution of this equation, we can
solve the original problem of eq (\ref{sch3}), using the coordinate transformations of  eq.(\ref{cond1}). 
For our problem, the transformed coordinate and the total potential are, after
restoring factors of $G$ and $c$,
\begin{equation}
z=\frac{2c}{3}\sqrt{\frac{2}{Gm_0}}a^{3/2}~, 
~~ \Omega(a) =\left(\frac{kac^4}{G}+\frac{7\hbar^2G}{64a^{3}c^2}-\frac{\Lambda c^4a^{3}}{3G}\right)~,~~
\Omega(z) =\frac{k}{2}\left(\frac{9m_0c^{10}}{G^2}\right)^{1/3}z^{2/3} +
\frac{7\hbar^2}{72 m_0 z^{2}} -
\frac{3}{8}c^2z^2m_0\Lambda~.
\label{Tr3}
\end{equation}
It is important to point out here that the $z^{-2}$ term is crucial in our analysis, and is often missed
in the literature. The broad reason is that the quantization procedure has to be done before making 
the transformation between the coordinates $z$ and $a$ in eq.(\ref{Tr3}), and {\it not} after \cite{AA}. For example,
had we simply made the transformation from the $a$ to the $z$ coordinate via the first relation 
in eq.(\ref{Tr3}), we would, from the second relation of eq.(\ref{Se1}) obtain a potential without the $z^{-2}$
piece, that vanishes at small values of $a$ (or $z$), which is clearly not physical. 
%In view of the importance of this issue,
%we have in an appendix, elaborated upon the details of the argument, following \cite{AA}. 

From eq.(\ref{Tr3}), we see that the behavior near the singularity $a=0$ i.e., $z=0$, is dominated by the $z^{-2}$ term, so that 
the wave function near the singularity is given as a linear combination of Bessel functions,  
and is hence regular. However, far from the singularity, the 
potential is dominated by the $z^{2}$ term. For a negative cosmological constant, the solution of the Schrodinger equation in this limit is that for
the simple harmonic oscillator.

Now, we will focus on the case $k=0$, i.e., marginally bound dust, and use the potential from eq.(\ref{Tr3}), which reads
\begin{equation}
\Omega(z) =\frac{7\hbar^2}{72m_0 z^{2}} - \frac{3}{8}c^2z^2m_0\Lambda~.
\label{ome3}
\end{equation}
For a negative cosmological constant, we can identify the potential of eq.(\ref{ome3}) as that of the 
isotonic oscillator, conventionally written in the form 
\begin{equation}
U_{iso}(z)=\frac{1}{2}m_{0}\omega^{2}z^{2}+ \frac{g}{2z^{2}}~, ~~ g\geq0~, 
\label{uiso}
\end{equation}
and in our case, $g =\frac{7\hbar^2}{36 m_0}, ~ \omega^{2}=-\frac{3c^2\Lambda}{4}$. In order
to avoid cluttering notation, we will henceforth remember that we are dealing only with the case of 
negative cosmological constant, and will write $\Lambda \equiv |\Lambda|$. 
The wave function and the energy eigenvalues of such oscillators are well studied (see, e.g. \cite{WJ}, \cite{CPRS}, \cite{is}). 
We will borrow the relevant results here. Denoting $\beta=\frac{1}{2}\sqrt{1+\frac{4m_0g}{\hbar^2}}$ (which, in
our case, reduces to $\beta = \frac{2}{3}$), 
the quantized energy values are given by
\begin{equation}
E_{n}=\left(2n+1+\beta\right)\hbar\omega, ~~~ n=0,1,2, \cdots
\label{isotonener}
\end{equation}
Since the potential is invariant under space inversion ($z\rightarrow-z$), the odd solutions have the same energy 
spectrum as that of the even solutions. The wave function for the even solutions in 
terms of the associated Laguerre polynomials $L^{\beta}_{n}(x)$ are given as \cite{is}
\begin{equation}
\phi_{n}(z)=\sqrt{\frac{2(m_0 \omega)^{1+\beta}n!}{\hbar^{1+\beta}\Gamma(n+\beta+1)}} z^{\frac{1}{2}+\beta}
\exp\left(-\frac{m_0 \omega z^2}{2\hbar}\right)L^{\beta}_{n}\left(\frac{m_0\omega z^{2}}{\hbar}\right), 
~~ \beta=\frac{2}{3}~,
\end{equation}
with the factor in front being the normalization constant. 
Finally, using the mapping in eq.(\ref{cond1}),  we can write down the normalized wave function in the original coordinate, after putting back 
the expression for $\omega$ as
\begin{equation}
\psi_{n}(a)=\sqrt{\frac{\Lambda^{\frac{1+\beta}{2}}n!}{\Gamma(n+\beta+1)}} \frac{2^{1+\beta}}{3^{\frac{1+3\beta}{4}}}
\frac{a^{1+\frac{3\beta}{2}}}{L_p^{1+\beta}}
\exp\left(-\frac{2a^3\sqrt{\Lambda}}{3\sqrt{3}L_p^2}\right)L^{\beta}_{n}\left(\frac{4a^3\sqrt{\Lambda}}{3\sqrt{3}L_p^2}\right)~,
~~ \beta=\frac{2}{3}~,
\label{wfn}
\end{equation}
where $L_p = \sqrt{\frac{\hbar G}{c^3}}$ is the Planck length.
Now, since the  associated Laguerre polynomials $L^{\beta}_{n}(x)$ have a constant value at $x=0$, the wave function 
goes to zero at $a\rightarrow 0$, indicating that the De Witt criterion for the  singularity avoidance has been satisfied.
Note also the emergence of the length scale $(L_p^2/\sqrt{\Lambda})^{1/3}$ associated with the collapsing
solution. 

\section{The quantum metric}
\label{secdBB}

So far we have considered the spacetime as classical, sourced by a classical distribution of the energy-momentum
tensor. Then we found out the quantum mechanical wave function of the collapsing system by using the 
functional Schrodinger formalism, but still assumed that the background remains classical, i.e., any quantum mechanical 
fluctuation of the energy momentum tensor is not taken into account. Now, if we neglect the backreaction due to Hawking 
radiation, this effect can be safely assumed to be  important only into the later stages of the collapse, where the density is 
so high that Planck scale physics comes into play. Incorporating these fluctuations in a general curved 
spacetime is an involved task, and is usually done by using the semi-classical Einstein equations, where one takes the expectation 
value of the energy-momentum tensor as the source of curvature, and also one has to renormalize this tensor 
suitably in the process \cite{Birrell}.  Our aim in this section will be to incorporate the quantum effects in the  
background classical metric in the case of a gravitational dust collapse model, by a method different from the above. 

Here, instead of following the standard method mentioned above, the approach 
we will use to incorporate the quantum effects in geometry is through the 
alternative dBB version of standard quantum mechanics. Here quantum 
particles have well defined trajectories (and hence positions and momenta), acted upon by a quantum potential term 
in addition to the usual external potential. By doing a polar decomposition of the  wave function and using it in 
the relevant wave equation (Schrodinger equation for non relativistic particle and Klein-Gordon equation for a 
relativistic one) we can determine the quantum potential, which is a nonlocal function of position and time \cite{PH}.

When one uses this approach to the motion of a particle  in a general curved background, 
the usual geodesic equation of the free particle becomes an acceleration equation, with an extra force coming 
from the quantum potential. In this context it is well known that the quantum effects can be systematically 
analyzed by geometrizing the problem, i.e., rescaling the metric itself by conformal transformation, 
with the conformal factor being a suitable function of the quantum potential. After doing this, 
the acceleration equation in the classical background becomes the force free geodesic equation of the 
transformed metric \cite{RC}. Since  the quantum effects are included  through the conformal factor, the  
transformed metric is said to be ``quantum corrected'' version of the original metric. 
Our goal here would be to do such a dBB analysis of the functional Schrodinger equation for the collapsing 
FRW dust system. Instead of a conformal transformation however, we will resort to a disformal transformation,
as we will elaborate shortly. 

\subsection{A de Broglie-Bohm interpretation of the functional Schrodinger equation. }

To do a dBB analysis of the functional Schrodinger eq.(\ref{sch1}), we start by  writing the wave function of eq.(\ref{sch3}) as
\begin{equation}\label{wave}
\Psi(a,\tau)=\mathcal{R}(a,\tau)e^{i\mathcal{S}(a,\tau)},
\end{equation}
where $\mathcal{R}(a,\tau)$ and $\mathcal{S}(a,\tau)$ are two single valued real functions of position and time 
\cite{PH}.\footnote{For a dBB interpretation of the minisuperspace version of 
the WDW equation see \cite{JV}, and in the context of FRW geometry see, e.g., \cite{YK}.} 
In the standard treatment of dBB theory, one substitutes this form of the wave function in the  
Schrodinger equation and after separating the real and imaginary  parts one gets the  Hamilton-Jacobi equation 
modified by the quantum potential term, and the  conservation equation of the  probability density 
$\rho=\mathcal{R}^{2}$. The quantum potential is obtained from 
$V_{qu}=-\frac{\hbar^{2}}{2m_{0}}\frac{\nabla^{2}\mathcal{R}}{\mathcal{R}}$, with $\nabla^{2}$ being the Laplacian. 
We also introduce velocity field 
along particle trajectory given by the formula $\mathbf{v}=d\mathbf{x}/dt=\hbar\mathbf{\nabla}\mathcal{S}/m_{0}$,  
with $m_{0}$ being the constant particle mass.

In applying this procedure to the present scenario, we first recall that for our system of collapsing homogeneous  
dust cloud, according to a comoving observer this equation is equivalent to the position dependent 
mass equation given in eq.(\ref{eq1}). Since  the procedure outlined above is applied 
only for a constant mass equation, the definition of the quantum potential has to be changed accordingly.  

Using eq.(\ref{wave}) in eq.(\ref{eq1}), separating real and 
imaginary parts, we get 
\begin{equation}\label{real}
-\hbar\bigg[\frac{\partial\mathcal{S}}{\partial \tau}\bigg]=\frac{\hbar^2}{2m(a)}\bigg[\frac{\partial\mathcal{S}}{\partial a}\bigg]^{2}
-\frac{\hbar^2}{2m(a)}\bigg[\frac{1}{\mathcal{R}}\frac{\partial^{2}\mathcal{R}}{\partial a^{2}}\bigg]+\frac{\hbar^2}{2\mathcal{R}m(a)^{2}}
\bigg[\frac{\partial m(a)}{\partial a}\bigg]\bigg[\frac{\partial \mathcal{R}}{\partial a}\bigg]+V(a)~,
\end{equation}
\begin{equation}\label{img}
\bigg[\frac{\partial \mathcal{R}^{2}}{\partial
\tau}\bigg]+\frac{\hbar}{m(a)}\frac{\partial}{\partial
a}\bigg[\mathcal{R}^{2}
\frac{\partial \mathcal{S}}{\partial
a}\bigg]=\frac{\hbar\mathcal{R}^{2}}{m(a)^{2}}\bigg[\frac{\partial
m}{\partial a}\bigg]
\bigg[\frac{\partial \mathcal{S}}{\partial a}\bigg].
\end{equation}
Note that the extra terms that arise due to the variable mass, namely the third term on the right hand side of eq.(\ref{real}), 
and the term on the right hand side of eq.(\ref{img}) will vanish if the mass is constant. 
These extra terms modify the quantum potential and the conservation equation. Now to interpret eq.(\ref{real}) as 
the quantum HJ equation, we introduce the following definitions of momenta, quantum Hamiltonian and the quantum potential 
respectively, 
\begin{equation}
p_{a}=\hbar\bigg[\frac{\partial \mathcal{S}}{\partial a}\bigg], ~~ \mathcal{H}_{qu}=-\hbar\bigg[\frac{\partial\mathcal{S}}
{\partial \tau}\bigg], ~~ V_{qu}(a)=-\frac{\hbar^2}{2m(a)}\bigg[\frac{1}{\mathcal{R}}\frac{\partial^{2}\mathcal{R}}
{\partial a^{2}}\bigg]+\frac{\hbar^2}{2\mathcal{R}m(a)^{2}}\bigg[\frac{\partial m(a)}{\partial a}\bigg]
\bigg[\frac{\partial \mathcal{R}}{\partial a}\bigg],
\label{phvq}
\end{equation}
so that it reduces to 
\begin{equation}
\mathcal{H}_{qu}=\frac{p_{a}^{2}}{2m(a)}+V(a)+V_{qu}(a)~~\equiv \mathcal{H}_{class}+V_{qu}~,
\label{QHJ}
\end{equation}
where the subscripts ``$qu$'' and ``$class$'' denote quantum and classical respectively. 
Now this represents motion with a variable mass $m(a)$ and a total potential $V+V_{qu}$. As usual, we can 
determine the quantum trajectories $a(\tau)$ by using the definition of the momenta, where  $\tau$ parameterize 
the quantum trajectory.

Now if we identify $p_{a}$ with the previous definition in eq.(\ref{sch3}) we get the unit vector along these quantum trajectories
\begin{equation}
\partial_{\tau}=\frac{\hbar}{2a}\big(\partial_{a}\mathcal{S}\big)\partial_{a}\equiv v\partial_{a}~,
\end{equation}  
where $v$ is the associated velocity $v=\frac{\hbar}{m(a)}\big(\partial_{a}\mathcal{S}\big)$.
However this parameter  is not related to any external clock and hence is not an observable (see \cite{JV} for related discussions). 
On the other hand, eq.(\ref{img}) can be interpreted as the continuity equation with the
velocity $v$ by introducing the probability density $\rho=\mathcal{R}^{2}(a)$ 
\begin{equation}
\bigg[\frac{\partial \rho}{\partial \tau}\bigg]+\frac{\partial}{\partial a}\big[\rho v\big]=0~.
\end{equation}

Now, to find out the quantum trajectories for the wave function we have obtained previously, let us suppose that 
at the start of collapse (taken at $\tau_{0}=0$ for convenience), the system is in a linear combination of the 
stationary state wave functions $\psi_{n}(a)$ obtained in eq.(\ref{wfn}). For simplicity we take this to be a linear 
superposition of the ground state ($n=0$) and the first exited state ($n=1$) wave functions, 
\begin{equation}
\Psi(a,0)=c_{0}\psi_{0}(a)+c_{1}\psi_{1}(a)~,
\end{equation} 
where $c_{0}, c_{1}$ are two constants which can be complex in general, and as can be seen form eq.(\ref{wfn}), 
both $\psi_{0}(a)$ and $\psi_{1}(a)$ are real functions of $a$. At any time $\tau>0$ during the collapse, this wave function 
evolves in the usual way,
\begin{equation}\label{nosta}
\Psi(a,\tau)=c_{0}\psi_{0}(a)e^{-iE_{0}\tau/\hbar}+c_{1}\psi_{1}(a)e^{-iE_{1}\tau/\hbar}~~= c_{0}\psi_{0}(a)
\big[e^{-iE_{0}\tau/\hbar}+dB(a)e^{-iE_{1}\tau/\hbar}\big]~,
\end{equation} 
with $E_{0}=\frac{5\hbar\omega}{3}, E_{1}=\frac{11\hbar\omega}{3}$ being the ground state and first exited state energies respectively
(remember that $\omega^2 = 3c^2\Lambda/4$ see discussion just after eq.(\ref{uiso})), 
from eq.(\ref{isotonener}) . To simplify the notation further, we will choose $c_0 = c_1 = 1/\sqrt{2}$, so that 
\begin{equation}
B(a)= L^{\frac{2}{3}}_{1}\left( \frac{4a^3\sqrt{\Lambda}}{3\sqrt{3}L_p^2}\right)~,~~ d =  \sqrt{\frac{\Gamma\left[\frac{5}{3}\right]}
{\Gamma\left[\frac{8}{3}\right]}}=\sqrt{\frac{3}{5}},
\label{bandd}
\end{equation}
Now from the polar form of eq.(\ref{wave}), we can glean the two real functions \cite{PH}
\begin{eqnarray}
\mathcal{R}(a,\tau)^{2}&=&|\Psi(a,\tau)|^{2}=\frac{1}{2}\left(\psi_{0}^{2}(a)+\psi_{1}^{2}(a)\right)+
\psi_{0}(a)\psi_{1}(a)\cos\left[\big(E_{0}-E_{1}\big)\frac{\tau}{\hbar}\right]~,\nonumber\\
\mathcal{S}(a,\tau)&=&\arctan\bigg[\bigg(\frac{1-dB(a)}{1+dB(a)}\bigg)\tan\big[\frac{1}{2}
\big(E_{0}-E_{1}\big)\frac{\tau}{\hbar}\big]\bigg]-\frac{1}{2}\big(E_{0}+E_{1}\big)\frac{\tau}{\hbar}~.
\label{polarform}
\end{eqnarray}
Note that the probability density $\mathcal{R}(a,\tau)^{2}=|\Psi(a,\tau)|^{2}$ is an oscillating function of time, 
and thus as expected, $\Psi(a,\tau)$ is not a stationary state.  

From the second relation in eq.(\ref{polarform}), we can calculate the momentum $p_a = \hbar\partial S/\partial a$ and
hence the velocity $v = \frac{1}{m(a)}p_{a}(a)$. We obtain
\begin{equation}
v =  \frac{18\sqrt{3}cd L_p^4\sqrt{\Lambda}a(\tau)\sin\left(c\sqrt{3\Lambda}\tau\right)}
{8\sqrt{3\Lambda}dL_p^2a(\tau)^3\left(5d+3\cos\left(c\sqrt{3\Lambda}\tau\right)\right)
-3L_p^4\left(9+25d^2+30d\cos\left(c\sqrt{3\Lambda}\tau\right)\right)-16d^2\Lambda a(\tau)^6}~.
\label{momfull}
\end{equation}
From this, we can solve for $a(\tau)$. It is difficult to obtain an analytic solution for $a(\tau)$ from eq.(\ref{momfull}). 
However, solving this equation numerically, we find that the quantum trajectories never reach $a(\tau)=0$, but 
shows bouncing behaviour. This is an indication that in the presence of quantum effects, the collapse does not reach the 
classical singularity. To substantiate this, it is useful to express the velocity in terms of the dimensionless
variable $l = a^3\sqrt{\Lambda}/L_p^2$. Now in the limit of small $l$, we obtain
\begin{equation}
v = -\frac{6\sqrt{3}cd\sqrt{\Lambda}a(\tau)\sin\left(c\sqrt{3\Lambda}\tau\right)}
{9 + 25d^2 + 30d\cos\left(c\sqrt{3\Lambda}\tau\right)}~,~{\rm i.e.,}~~
a(\tau)=\mathcal{C}_1\bigg[9+25d^2 + 30d\cos\left(c\sqrt{3\Lambda}\tau\right)\bigg]^{1/5}
\label{ata}
\end{equation}
with $\mathcal{C}_1>0$ being an integration constant, which can be determined from the initial value of the scale factor.
It can be checked that for small values of $\mathcal{C}_1\lesssim 0.1$, the second relation of eq.(\ref{ata}) gives an excellent
approximation to the full solution of the scale factor obtained from eq.(\ref{momfull}), for a given initial value of the scale factor. 
Also, in terms of $l$, the quantum potential computed 
from the third relation of eq.(\ref{phvq}) via the definition of ${\mathcal R}$ in the first equation of eq.(\ref{polarform}) yields 
in the limit of small $l$, the simple expression
\begin{equation}
V_{qu}=\frac{c\hbar}{2}\sqrt{\frac{\Lambda}{3}}\left(8 + \frac{3\left(25d^{2}-9\right)}{25d^{2}+9+30d \cos\left(c\sqrt{3\Lambda}\tau\right)}\right)
=\frac{c\hbar}{2}\sqrt{\frac{\Lambda}{3}}\left(8 +18 \left(\frac{\mathcal{C}_{1}}{a(\tau)}\right)^{5}\right)~,
\label{qpot}
\end{equation}
where the second relation follows from eq.(\ref{ata}). From eqs.(\ref{ata}) and (\ref{qpot}) we see that indeed the quantum 
potential maximizes when the position of the particle is closest to the classical singularity \cite{BNL}. 

In fig.(\ref{fig:scale-factor}), we show this graphically. Here the scale factor (solid red) and the quantum 
potential (dotted blue) are plotted as a function of $\tau$, in the limit of small $l$. 
In the graph, the potential has been scaled by a factor of $10^{-2}$ to offer a comparison with $a(\tau)$, and we
have set $c = \hbar = \Lambda = 1$, with ${\mathcal C}_1=10^{-2}$.
As can be seen from fig.(\ref{fig:scale-factor}), when the position of the particle is close to the classical singularity, 
the effect of the quantum potential is maximum, and this is the reason for the bouncing behavior. On the other hand, 
when the shell is at its maximum radius (away from the singularity), the quantum potential is at its minimum. 
With our chosen constants, the bounces occur at $\tau = n\pi/\sqrt{3}$, with integer $n$. 

\subsection{The quantum collapsing solution}

Now that we have the  characterization of the particle position, we want to find out a quantum version of the FRW 
solution.  To understand what is meant by the the quantum version of the metric we start by writing the 
$\mathcal{H}_{qu}$ of eq.(\ref{QHJ})  in terms of the mass function obtained from the original metric $g_{\mu\nu}$, 
using eq.(\ref{MisnerSharp}), as
\begin{equation}
\mathcal{H}_{qu}= \mathcal{H}_{class}+V_{qu}~\equiv  \frac{M(r,a)c^2}{r^{3}}+\frac{1}{3}\frac{c^4}{G}\Lambda a^{3}+V_{qu}(a)~.
\label{QuHa}
\end{equation}
We will call the metric $\bar{g}_{\mu\nu}$  the quantum corrected solution when the  dynamics of the collapse 
with respect to this metric will give rise to the following conserved quantity as it's classical Hamiltonian
\begin{equation}\label{ClHa}
\bar{\mathcal{H}}_{class}= \frac{\bar{M}(r, A(t))c^2}{r^{3}}+\frac{1}{3}\frac{c^4}{G}\Lambda A^{3}+V_{qu}(A)= Constant,
\end{equation}
where $A(t)$ denotes the new scale factor and $\bar{M}(r, A(t))$ is the new mass function constructed out of 
the metric $\bar{g}_{\mu\nu}$. We have denoted this barred Hamiltonian 
by a subscript to emphasize that this represents the motion of a classical particle in the quantum background. 
Thus here the quantum potential term $V_{qu}$ is not due to the quantum nature of the particle (as it was in eq.(\ref{QuHa})),  
instead it must come from the part of  the  energy momentum  tensor created by the quantum effects. 
Thus the original quantum effects in the collapsing classical background is equivalent to this  classical motion in the 
background of such a ``quantum metric'' $\bar{g}_{\mu\nu}$. The advantage of the new quantum version of the metric 
is that geometric quantities computed from here will give a clearer picture 
of the collapsing solution. For this, we have to choose a suitable ansatz. 

If we assume a completely general diagonal  form for the metric as a possible solution, it will have at least three 
unknown function of spacetime coordinates, and the single requirement that its mass function gives the Hamiltonian of eq.(\ref{ClHa}) is not 
enough for determining all of them. So, we will work with a less general solution and 
make the following assumptions for the quantum 
metric : (1) It is of the form of a FRW solution with a new scale factor $A(\tau)$. 
(2) One can obtain this new solution from the classical one by using a conformal or disformal transformation 
(see below), with the transformation factors carrying the quantum effects through the quantum potential. (3) 
The solution is free from the classical 
singularity at the zeros of the scale factor. The last two points needs further clarification, and we start 
with the singularity resolution criteria.

The wave functions we have obtained vanish at the classical singularity at $a=0$ and, according to De Witt criterion, this is a
sufficient condition for the singularity avoidance \cite{gotay}. But the De Witt criterion does not guarantee the  quantum corrected 
solution will be singularity free. One can perfectly well construct a singular solution which will satisfy the quantum HJ equation. 
Furthermore one can encounter solutions which are singularity free at $a(\tau)=0$ (corresponding to, say $\tau=\tau_{s}$), 
but which still have a singularity at time $\tau=\tau_{s1}>\tau_{s}$ which corresponds to a zero of the new solution i.e.,
$A(\tau)=0$, and since we want the solution to be singularity free for all $\tau>0$, we will not consider such cases. 
Note that since  we can always rescale the time coordinate of the FRW metric by changing the lapse function (we call it $N(t)$) to
define a new coordinate $t$, one might argue that the last kind of solutions can be made singularity free by 
changing the definition of  time. However, if the lapse function $N(t)$ has a zero at finite $t$, then such a transformation 
will not be globally defined. 

Now we will explain the second criterion, i.e., the new solution being related to  the classical one by a conformal or a 
disformal transformation. It is well known that in the dBB version of quantum mechanics, one can transform to 
a conformal frame, with the conformal factor being a suitable function of the quantum potential, such that  the new 
metric contains all the quantum effects (see, e.g. \cite{RC}). If the original metric is  
singular, then in the transformed frame, the singularity shows up as the zeros of the conformal factor, 
making the inverse transformation undefined at the singularity. But in this procedure, one faces a few well known problems.
For example, in the transformed frame, one does not get the correct continuity equation, the massless particles do 
not have the correct description, and most  importantly, the problem of a negative conformal factor might arise. 
It can be shown that \cite{QM} one can avoid  these problems by 
demanding that the new solution $\bar{g}_{\mu\nu}$ be related to $g_{\mu\nu}$ by a general disformal  transformation of the type
\begin{equation}
\bar{g}_{\mu\nu}=\Theta^{2}(\phi,X)g_{\mu\nu}-\epsilon \mathcal{B}(\phi, X)\phi_{\mu}\phi_{\nu}, ~~ \bar{g}^{\mu\nu}=
\Theta^{-2}\bigg[g^{\mu\nu}+\epsilon \frac{\mathcal{B}}{\Theta^{2}-2X\mathcal{B}}\phi^{\mu}\phi^{\nu}\bigg]~,~~~ 
X=-\frac{1}{2}g^{\mu\nu}\phi_{\mu}\phi_{\nu}~.
\label{Dist}
\end{equation}         
Here  $\Theta$ and $\mathcal{B}$ are arbitrary real functions of a scalar field $\phi$, and 
$\phi_{\mu}=\nabla_{\mu}\phi$ is the 
normal vector to the $\phi=$ constant hypersurface. The nature of this vector is  determined by the value of $\epsilon$, 
which is $1,0,-1$ for timelike, null and spacelike vectors, respectively. 
%%%%%%%%%%%%%%%%%%%%%%%%
\begin{figure}[t!]
	\begin{minipage}[b]{0.5\linewidth}
		\centering
		\includegraphics[width=2.4in,height=1.8in]{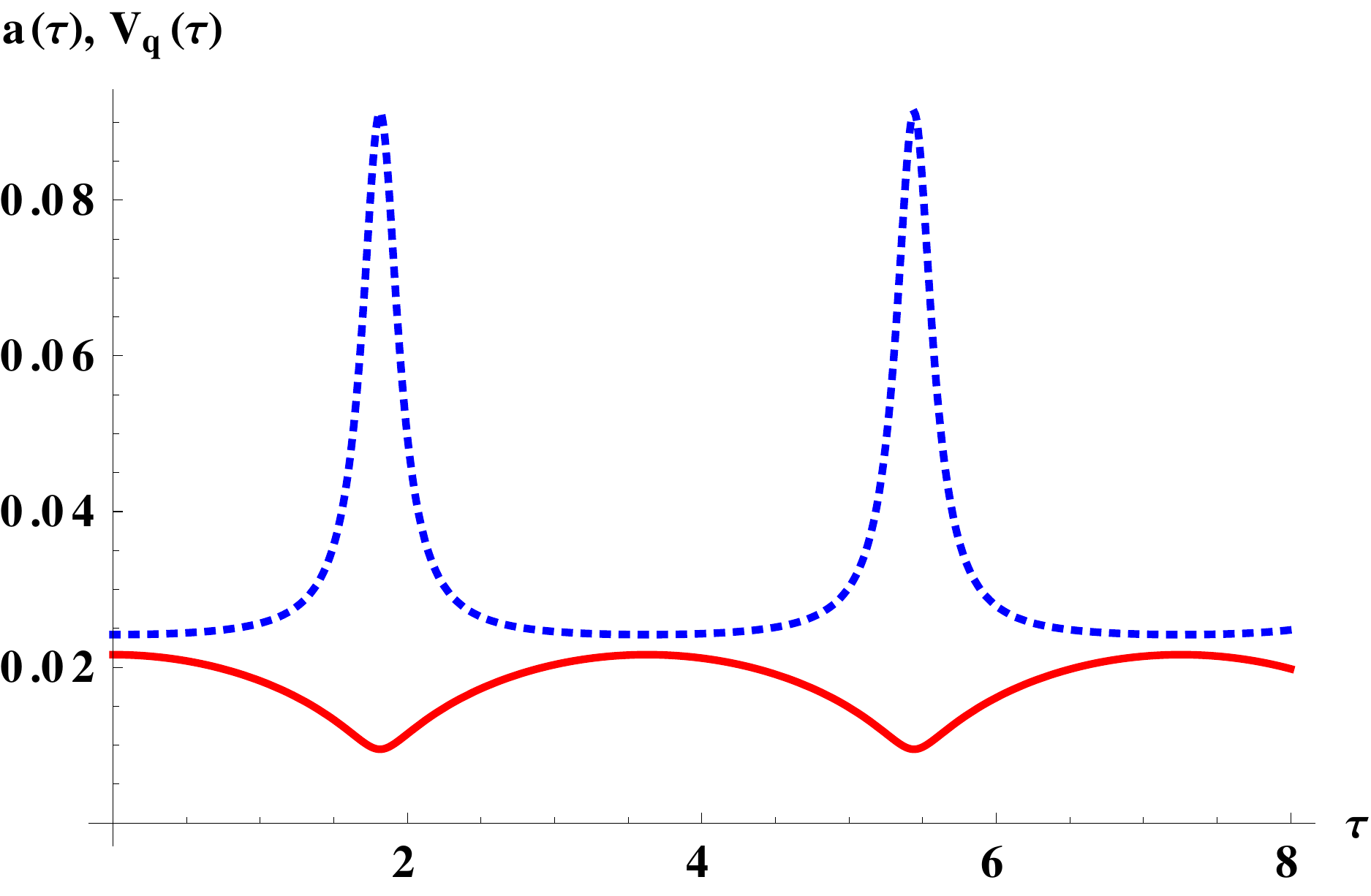}
\caption{Evolution of scale factor (red) and the quantum potential (dotted blue)  with time. }
\label{fig:scale-factor}
	\end{minipage}
    \hspace{0.2cm}
		\begin{minipage}[b]{0.5\linewidth}
		\centering
		\includegraphics[width=2.4in,height=1.8in]{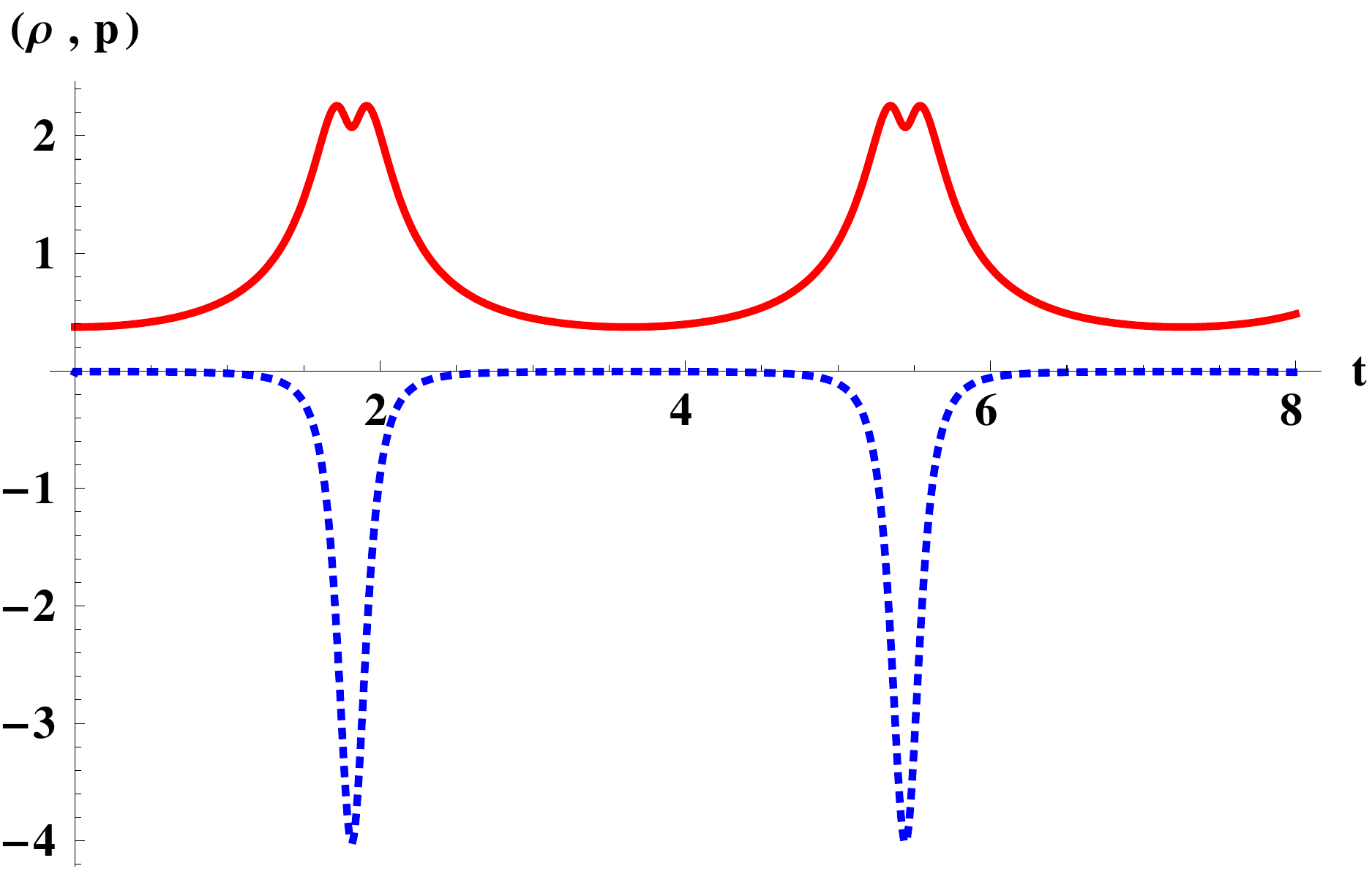}
\caption{Evolution of energy density (red) and pressure (dotted blue) of the interior metric with time.}
\label{denpr}
	  \end{minipage}
\end{figure}
%%%%%%%%%%%%%%%%%%%%%%

Here we consider the simplest case of pure disformal transformation, namely an anisotropic change of 
local geometry at every point along a particular direction chosen by the disformal vector,  where the conformal factor 
is a constant, which we shall take to be unity.
This also means that the scale factor is taken to be the same as the solution 
in eq.(\ref{ata}) of the equation of motion , i.e., $A(t)\equiv a(\tau=t)$.\footnote{We have denoted 
the time coordinate as $t$ for convenience, as $\tau$ usually denotes the proper time of the comoving observer. 
But note that when doing conformal and disformal transformations, the coordinate patch used to write down both the 
metrics must be same, and for that we will denote the $\tau$ used for classical solution as $t$ in this section.}  
In that case, we take the normal vector to be a timelike vector  pointing along the direction specified by the normal 
to the  $A(t)=$ constant hypersurface. Hence the transformed line element can be written as the sum of FRW metric 
(with $k=0$) and a pure disformal part
\begin{equation}
\bar{ds}_-^{2}={\bar g}_{\mu\nu}dx^{\mu}dx^{\nu}=-c^2dt^{2}+A(t)^{2}\Big[dr^2+r^{2}d\Omega^{2}\Big]+\mathcal{B}\dot{A}^{2}dt^{2}~~= 
-N(t)^{2}c^2dt^{2}+A(t)^{2}\Big[dr^{2}+r^{2}d\Omega^{2}\Big]~,
\label{QFRW}
\end{equation}
where $N(t) = (1 - \frac{{\mathcal B}{\dot A}^2}{c^2})$, 
and an overdot represents derivative with respect to $t$. Thus we  have only one arbitrary function 
$N(t)$ which can be determined as follows. 
With the metric of eq.(\ref{QFRW}), the Misner-Sharp mass is, from eq.(\ref{MisnerSharp}),
\begin{equation}
{\bar M}(r,A) = \frac{r^3A{\dot A}^2}{N^2G}~.
\label{MFq}
\end{equation}
In terms of the mass function ${\bar M}$, the
Einstein's equations are given by the analog of eq.(\ref{Eingen}), i.e., with a subscript $mat$ denoting
the matter part of the stress tensor as before, we have 
\begin{equation}
\frac{{\bar M}'}{Q^2Q'} = \rho_{mat}  + \rho_{qu} - \frac{\Lambda c^2}{G}~,
~~\frac{{\dot {\bar M}}}{Q^2{\dot Q}} =  -\frac{p_{qu}}{c^2} - \frac{\Lambda c^2}{G}~,
\label{EingenQ}
\end{equation}
where $Q(r,t) = rA(t)$, and 
$\rho_{qu}$, $p_{qu}$ are a priori unknown functions that arise entirely due to quantum corrections, and vanish when $V_{qu}$ is set to zero. 
Note that the quantum part of the density is given as 
\begin{equation}
\rho_{qu} = -\frac{3V_{qu}}{c^2 A(t)^3}~,
\label{rhoqu}
\end{equation}
which is negative definite and hence cannot arise due to a matter distribution. 
We know however that upon following the same procedure that led to 
eq.(\ref{conserved}), we should now get a conserved quantity that gives the Hamiltonian of eq.(\ref{ClHa}). 
If we integrate the second relation of eq.(\ref{EingenQ}), we will obtain a conserved quantity (denoted as $\mathcal{E}$) corresponding
to the Hamiltonian of eq.(\ref{ClHa})
\begin{equation}
\frac{{\bar Mc^2}}{r^3} + \frac{\Lambda c^4}{3G} A^3 + V_{qu} = {\mathcal E} ~,
\end{equation}
provided that  we identify the quantity
\begin{equation}
p_{qu}=\frac{\dot{V}_{qu}(t)}{A^{2}\dot{A}}~
\label{MFE}
\end{equation}
as the pressure created due to the quantum potential. Substituting the value of ${\bar M}$ from eq.(\ref{MFq}), we finally obtain 
\begin{equation}
\frac{c^2}{G}\frac{A(t){\dot A(t)}^2}{N^{2}(t)}+\frac{\Lambda c^4}{3G}A^{3}+V_{qu}(A)=\mathcal{E}~.
\label{coen}
\end{equation}
This constant of motion $\mathcal{E}$ is the classical Hamiltonian in eq.(\ref{ClHa}). Now  inverting the above relation, we 
get  the unknown function $N(t)$ to be
\begin{equation}
N(t)=\frac{c}{\sqrt G}\frac{\sqrt{A}\dot{A}}{\big[\mathcal{E}-V_{qu}(A)-\frac{\Lambda c^4}{3G}A^{3}\big]^{1/2}}~,
\label{nt}
\end{equation}
and this also gives the required  disformal factor $\mathcal{B}(t)$. 
With our earlier assertion that the scale factor $A(t) = a(\tau = t)$, using the wave function of eq.(\ref{wfn}) 
and further using the definition 
of $V_{qu}$ given in eq.(\ref{phvq}), we have a solution for $N(t)$ and hence for the metric of eq.(\ref{QFRW}), and
these reveal useful information. For example, with $G=c=1$, the Ricci scalar for the metric in eq.(\ref{QFRW}) is given as
\begin{equation}
R_-(t)=\frac{3\mathcal{E}-4\Lambda A^{3}-3V_{qu}}{A^3}-\frac{3\dot{V}_{qu}}{A^{2}\dot{A}}~,
\label{Rsca}
\end{equation}  
which is clearly non singular if $A\neq0$, as follows from eq.(\ref{qpot}). 

Before constructing a solution for the  nonstationary state of eq.(\ref{nosta}), we mention briefly what will happen 
when the initial state of the system is  a stationary state $\psi_{n}(a)$. Then using eqs.(\ref{phvq}) and (\ref{wfn}), 
we calculate the quantum potential to be of the form $V_{qu}(A) \sim \mathcal{N}_{n}-\frac{\Lambda A^{3}}{3}$, 
where the first term is a constant, which depends on the quantum number $n$.
Substituting this in eq.(\ref{coen}), it can be seen that, as expected for a stationary state, the quantum potential  
gives rise to a  pressure that is opposite to the cosmological constant. Here the quantum 
corrected system acts as if it is asymptotically flat space ($\Lambda=0$), and as is well known for dust 
collapse in flat background, the collapse always reaches the singularity. In this sense we had mentioned earlier 
that even if the initial wave function satisfies the De Witt criterion, the quantum metric can be singular. 
The nonsingular final state of collapse  depends on the initial state of the wave function.  

Now for the linear superposition of stationary states consider in the previous section, 
we already have the scale factor $A(t)$ from eq.(\ref{ata}), the disformal factor $N(t)$ from eq.(\ref{nt}) and the quantum 
potential from eq.(\ref{qpot}). 
Substituting these in eqs.(\ref{rhoqu}) and (\ref{MFE}), we get the expressions for the density and pressure 
for our model. In fig.(\ref{denpr}), with ${\mathcal C}_1=10^{-2}$ in eq.(\ref{ata}) as before, 
and setting $c = \hbar = G = \Lambda = 1$, we have plotted 
$\rho = \rho_{mat} + \rho_{qu}$ (solid red) and $p_{qu}$ (dotted blue) with ${\mathcal E} = 15$ as an illustration. 
The matter distribution with the quantum correction is that of a perfect fluid. 

The role of the quantum potential in avoiding the classical singularity should now be clear from the simple model 
we have constructed above. From fig.(\ref{denpr}), we see that the homogeneous pressure, coming from 
solely the quantum effects is small at the start of the collapse, 
but in the region where the shell bounces back, this increases dramatically, 
and its negative value  indicates that it originates from some exotic source other than classical matter distribution.\footnote{Such 
bouncing behavior of a collapsing  matter distribution is common in a semiclassical treatment, where one attribute 
such behavior to the modified density and pressure coming from quantum corrections, which becomes important at the 
later stage of the collapse  (see \cite{Malafarina} for a review).}  

We can check whether the classical energy conditions are satisfied by this matter at the start and  during the collapse process. 
With the numerical values chosen for fig.(\ref{denpr}), it can be checked that the weak, null and dominant energy conditions are 
violated close to the bounce at $t = n\pi/\sqrt{3}$. However, for a sufficiently large value of 
${\mathcal E}$, both conditions can be satisfied. The violation of a classical
energy condition is not unexpected in a semi-classical treatment such as ours. 
It is well known that in a semi-classical theory of gravity, all or some of 
the classical energy conditions can indeed be violated  \cite{BMM}, \cite{Visser1}, \cite{Visser2}. 

We now discuss the matching of our interior quantum collapsing solution with an exterior spacetime. 
Such a matching is performed at a timelike hypersurface $\Sigma$  characterized by the constant value 
of the radial coordinate $r=r_{0}$. The matching at $\Sigma$ is said to smooth if the Israel junction conditions
$\big[h_{ab}\big]=0,~~ \big[K_{ab}\big]=0$
are satisfied at the junction  \cite{EP}. Here $h_{ab}$ and $K_{ab}$ are the induced metric on the hypersurface, 
and the extrinsic curvature respectively, and the notation $ \big[X\big]$ implies the change of the 
quantity $X$ across the junction.

As we have mentioned before, in the presence of the quantum effects taken through the quantum potential 
$V_{q}(A)$, the energy momentum tensor of the matter distribution behaves as a perfect fluid. As is  well known,
in such cases, the interior FRW metric can be matched smoothly with an exterior Vaidya solution\cite{GJ,RWS}. 
This can be straightforwardly done for our case to match the FRW like solution of 
eq.(\ref{QFRW}) with an AdS Vaidya solution.

However, it is more reasonable here to consider a slightly different situation. Namely that, since the interior 
metric has been quantum corrected through the quantum potential, we can envisage an exterior static solution with
quantum correction terms, with these terms being determined again through the quantum potential. This
must then have a non-zero energy momentum tensor arising out of quantum corrections.\footnote{Any vacuum solution with negative cosmological 
constant should be SAdS, according to the Birkhoff's theorem.} 
In this spirit, let us consider the following ansatz for the quantum corrected Schwarzschild AdS solution in 
$(\mathbf{t},\mathbf{r},\theta, \phi)$ coordinates\footnote{We will
use bold cases for the coordinates of the external metric.} 
\begin{equation}
ds^{2}_{+}=-c^2f_{q} d\mathbf{t}^{2}+f_{q}^{-1} d\mathbf{r}^{2}+\mathbf{r}^{2} d\Omega^{2}~,~~ f_{q}(\mathbf{r})
=1-\frac{2G\mathcal{M}}{c^{2}\mathbf{r}}+\frac{\Lambda}{3}\mathbf{r}^{2}+\mathcal{V}_{q}(\mathbf{r})~,
\label{qextirior}
\end{equation}
with the quantum correction term $\mathcal{V}_{q}(\mathbf{r})$,
as the exterior to the collapsing dust sphere. This is expected, since quantum effects are taken in the metric itself,
and as was the case in the interior, the exterior metric should also corresponds to some non zero 
matter distribution due to $\mathcal{V}_{q}(\mathbf{r})$.
We want to determine $\mathcal{V}_{q}(\mathbf{r})$ 
by demanding that the external solution be matched with the quantum corrected interior of eq.(\ref{QFRW})  
through a timelike hypersurface $r=r_{0}$. The procedure is standard.

We start by writing eq.(\ref{QFRW}) in a form analogous to the ordinary FRW metric,
\begin{equation}
\bar{ds}_{-}^{2}= 
-c^{2} d\tau^{2}+A(\tau)^{2}\Big[dr^{2}+r^{2} d\Omega^{2}\Big]~,~~d\tau=N(t)dt~,
\label{QFR2W}
\end{equation}
where $\tau$ is the proper time of the comoving observer. As seen from outside of the collapsing sphere, 
the hypersurface is determined by the parametric relations $\mathbf{r}= \mathbf{r}_{0}(\tau)$ 
and $\mathbf{t}= \mathbf{t}_{0}(\tau)$. 
Then the first junction condition, namely the continuity of the induced metric implies the following relations 
\begin{equation}\label{istjc}
\mathbf{r}_{0}=r_{0}A(t)~,~~ c^2\big(\mathbf{t}_{0\tau}\big)^{2}F_{q}-\big(\mathbf{r}_{0\tau}\big)^{2}
F_{q}^{-1}=c^2~,~~ \text{with}~~
F_{q}=f_{q}(\mathbf{r}=\mathbf{r}_{0})~.
\end{equation}
Here the subscript $\tau$ indicates derivative with respect to proper time.  The 
continuity of the $tt$ and $\theta\theta$ components of the extrinsic curvature can be shown to imply that
$\sqrt{c^2F_{q}+\mathbf{r}_{0\tau}^{2}}=c$.

Then, using the first relation of eq.(\ref{istjc}), and finally restoring the comoving time 
$t$, we get 
\begin{equation}
\mathcal{M}=\frac{c^2Q_{0}}{2G}\Big(\frac{\dot{Q_{0}}^{2}}{c^2N^{2}(t)}+\frac{\Lambda}{3}Q_{0}^{2}
+\mathcal{V}_{q}(Q_{0})\Big)~,~~ \text{with} ~~Q_{0}=r_{0}A(t)~.
\label{qmet1}
\end{equation}
Now we compare this equation with the conserved classical Hamiltonian of the quantum metric given 
in eq.(\ref{coen}) evaluated at the comoving boundary $Q_{0}$. From 
eqs.(\ref{qpot}) and eq.(\ref{coen}), we have
\begin{equation}
\frac{c^{2}}{G}\left[\frac{A(t)}{N^{2}(t)}\dot{A}^{2}+\frac{c^{2}\Lambda}{3}A^{3}+\frac{9G\hbar}{c}
\sqrt{\frac{\Lambda}{3}} \left(\frac{\mathcal{C}_{1}}{A(t)}\right)^{5}\right]=\mathcal{E}-4c\hbar\sqrt{\frac{\Lambda}{3}}~.
\label{qmet2}
\end{equation}
As can be seen by comparing eqs.(\ref{qmet1}) and (\ref{qmet2}), if we set 
\begin{equation}
\mathcal{V}_{q}(\mathbf{r})=\frac{G\hbar}{c^{2}}\frac{r_{0}^{8}\mathcal{C}_{2}}{\mathbf{r}^{6}}
=L_{p}^{2}\frac{r_{0}^{8}\mathcal{C}_{2}}{\mathbf{r}^{6}}~,~~\text{with}~~\mathcal{C}_{2}=
9\sqrt{\frac{\Lambda}{3}}\mathcal{C}_{1}^{5}~~,
\label{C1C2}
\end{equation}
then the mass $\mathcal{M}$ is constant and  is completely determined in terms of the constant of 
motion $\mathcal{E}$ and $r_{0}$ by 
\begin{equation}
\mathcal{M}=\frac{r_{0}^{3}}{2c^{2}}\Big(\mathcal{E}-4c\hbar\sqrt{\frac{\Lambda}{3}}\Big)~.
\label{MExp}
\end{equation}
This completes our task of smooth matching of the quantum interior with a quantum corrected exterior solution.  
The quantum corrected metric is of the form eq.(\ref{qextirior}) with $ f_{q}(\mathbf{r})
=1-\frac{2G\mathcal{M}}{c^{2}\mathbf{r}}+\frac{\Lambda}{3}\mathbf{r}^{2}+\frac{L_{p}^{2}r_0^8\mathcal{C}_{2}}{\mathbf{r}^{6}}$.

The external metric of eq.(\ref{qextirior}) can now be written in terms of $\mathcal{M}$ as 
\begin{equation}
ds_+^2 = -c^2 f_{q}(\mathbf{r})d{\mathbf t}^2 + f_{q}(\mathbf{r})^{-1}d{\mathbf r}^2+ {\mathbf r}^2d\Omega^2~,~{\rm with},~
f_{q}(\mathbf{r})=1-\frac{2G\mathcal{M}}{c^{2}\mathbf{r}}+\frac{\Lambda}{3}\mathbf{r}^{2}+\frac{L_{p}^{2}\mathcal{C}_{2}}
{\mathbf{r}^{6}}\Bigg[\frac{2\mathcal{M}c^{2}}{\mathcal{E}-4c\hbar\sqrt{\frac{\Lambda}{3}}}\Bigg]^{8/3}~.
\label{ext}
\end{equation}  
We point out that a quantum correction term in the Schwarzschild solution proportional to $1/{\mathbf r}^{6}$ is 
known in the literature. In \cite {CM}, the authors have shown using the techniques of effective field theory 
applied to GR, when the effective  Lagrangian contains correction in terms of operators  of dimension $\geq6$, the 
standard Schwarzschild metric does not satisfy the GR field equation and is modified by a correction term proportional 
to $1/{\mathbf r}^{6}$ to the lowest order. However generalization of such calculations when a non zero cosmological constant 
is present is a non trivial task. Importantly, the correction found in \cite{CM} is proportional to the square of the
Schwarzschild mass. In this case, the quantum correction is proportional to $\mathcal{M}^{8/3}$, indicating its
difference from that obtained in an effective field theory. 

The exterior static solution that we have constructed must have non-zero principal pressures. 
The matter content of the external metric of eq.(\ref{ext}) is gleaned by computing its energy momentum tensor. We find
\begin {equation}
\rho = -p_r = \frac{5K}{{\mathbf r}^8}~,~~p_{\theta}=p_{\phi}=3\rho~,~{\rm where}~~
K = \frac{5{\mathcal C}_2\hbar r_0^8}{8\pi c}~ = \frac{5{\mathcal C}_2\hbar}{8\pi c}\left(
\frac{2{\mathcal M}c^2}{{\mathcal E} -4c\hbar\sqrt{\Lambda/3}}\right)^{8/3}~.
\label{emtext}
\end{equation}
This represents an anisotropic fluid whose density and principal pressures arise entirely from quantum effects. 
As expected, these will be negligible at large values of ${\mathbf r}$. 
A static, anisotropic fluid is known to have an energy momentum tensor of 
the form $\rho = -p_r,~p_{\theta} = p_{\phi}=\alpha\rho$, where $\alpha$ is a constant. To satisfy all the 
classical energy conditions, we need  $\rho \geq 0$ and $0\leq \alpha \leq 1$. In the quantum case, we see that the 
dominant energy condition is not satisfied, while the weak, strong and null energy conditions are. 
Again, this is not unexpected as we have mentioned before, as the violations are purely due to quantum effects. 
We also record the expression for the Ricci scalar, given by
\begin{equation}
R_+ = -\frac{20{\mathcal C}_2L_p^2r_0^8}{r^8} - 4\Lambda~.
\label{Ric}
\end{equation}
This indicates that the external space-time is no longer of constant curvature, with the departure being entirely
quantum in nature. 

Before ending this section, we point out that bounds on the quantum correction term in the external metric
can be estimated from solar system constraints. 
In order to show this, we consider the redshift factor
$\nu_A/\nu_B = \sqrt{g_{{\mathbf t}{\mathbf t}({\mathbf r}_A)}/g_{{\mathbf t}{\mathbf t}({\mathbf r}_B)}}$, with
${\mathbf r}_A$ and ${\mathbf r}_B$ denoting two different radii. As an illustration, we substitute ${\mathcal E}=1~{\rm TeV}$ 
and the earth's mass for ${\mathcal M}$ in eq.(\ref{ext}). Following \cite{Kagramanova},
we assume that clock comparison is done with an accuracy of $10^{-15}$, and consider such a comparison
between the earth's surface and a satellite $15\times 10^3$ Km above this surface. Then with an assumed value
of $\Lambda \sim 10^{-52}~{\rm meter}^{-2}$ , we obtain the upper bound ${\mathcal C}_1 \lesssim 0.015~{\rm meters}$.

\section{Horizon structure}
\label{Hor}

\begin{figure}[h!]
\begin{minipage}[b]{0.3\linewidth}
\centering
\includegraphics[width=2.4in,height=1.8in]{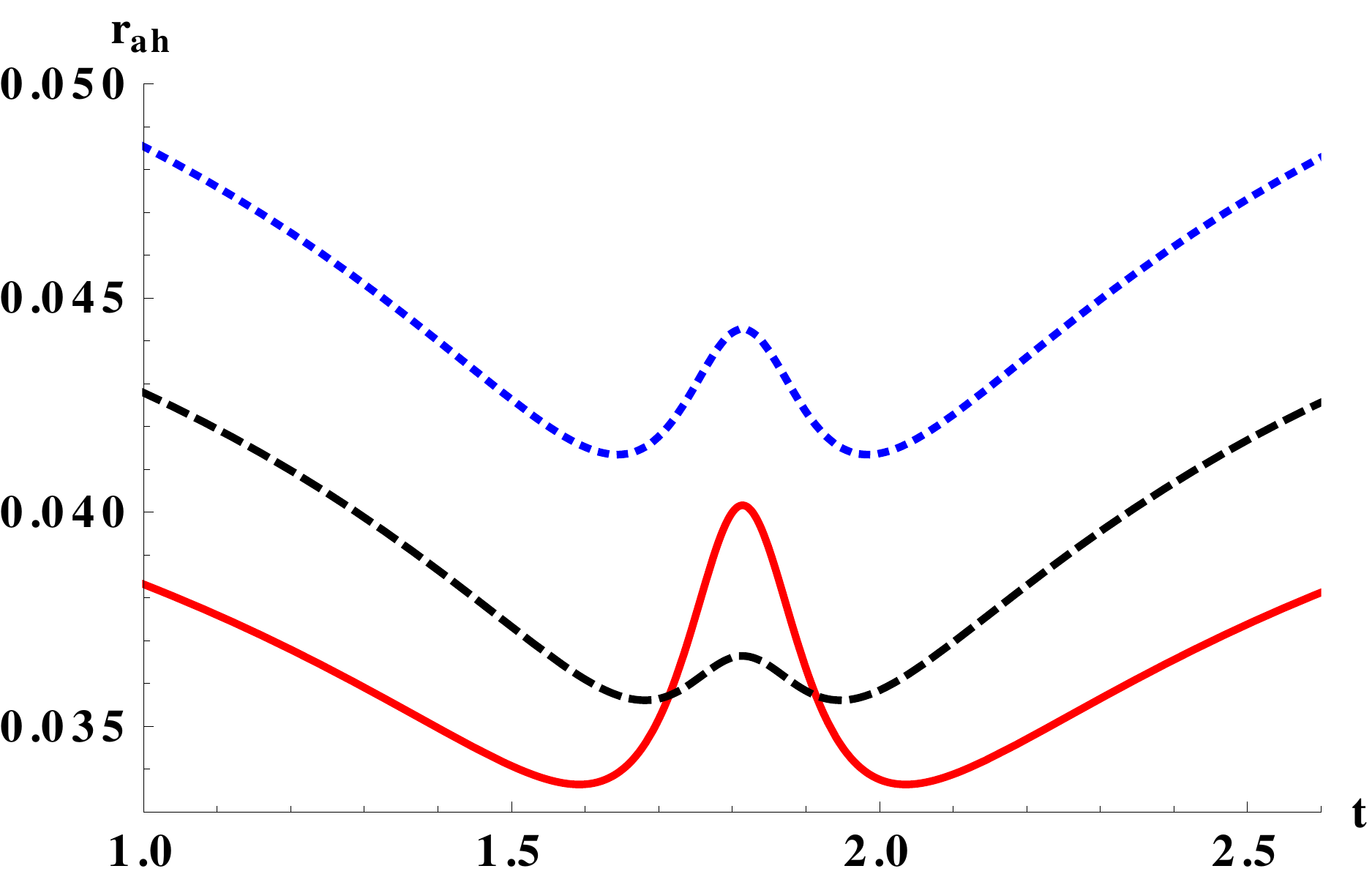}
\caption{Apparent horizon curves for different values of $\mathcal{E}$. }
\label{fig:app horizons}
\end{minipage}
\hspace{0.2cm}
\begin{minipage}[b]{0.3\linewidth}
\centering
\includegraphics[width=2.4in,height=1.8in]{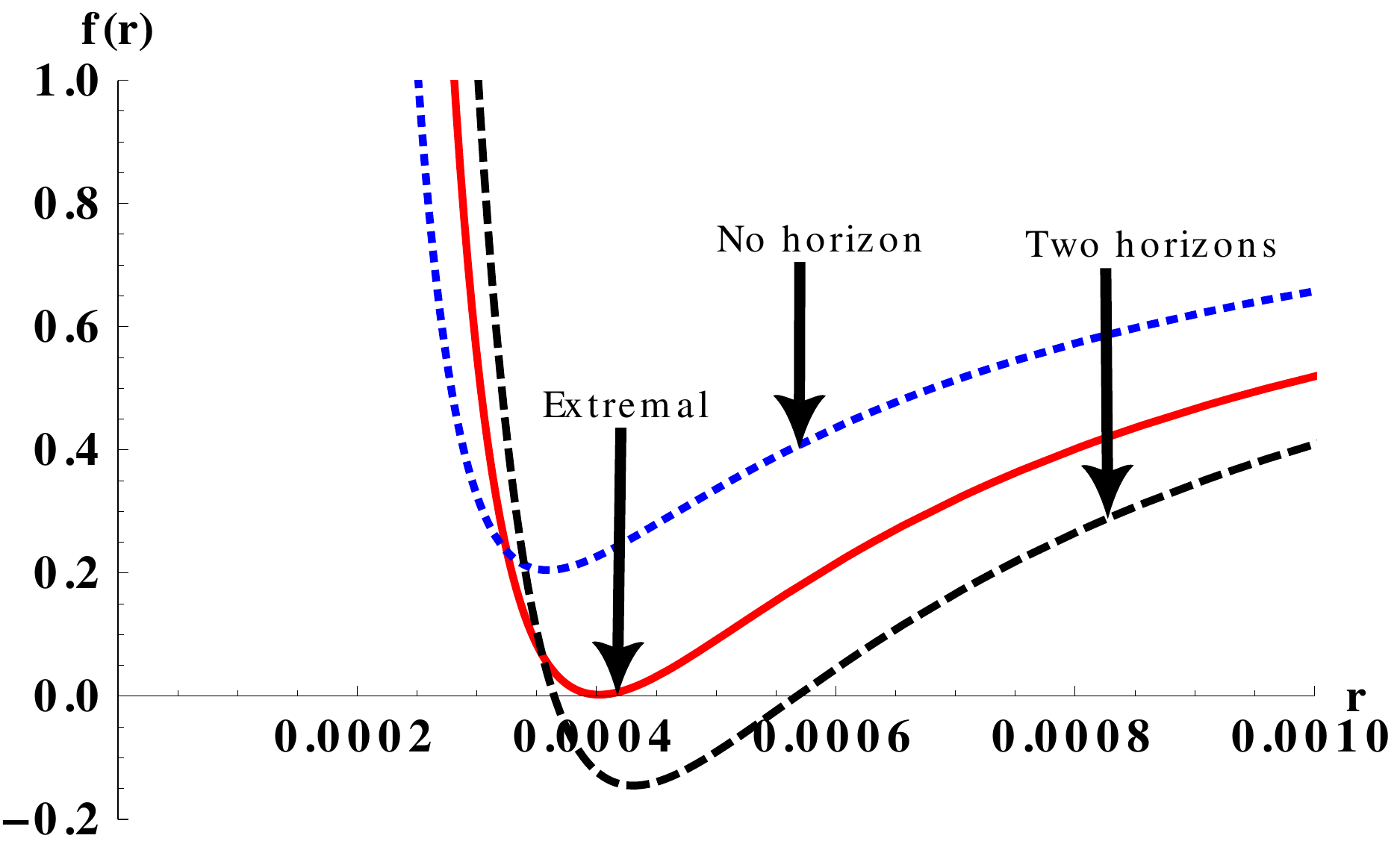}
\caption{Plot of $f(\mathbf{r})$ for different values of $r_{0}$ }
\label{fig:inner and outer}
\end{minipage}
\hspace{0.2cm}
\begin{minipage}[b]{0.3\linewidth}
\centering
\includegraphics[width=2.4in,height=1.8in]{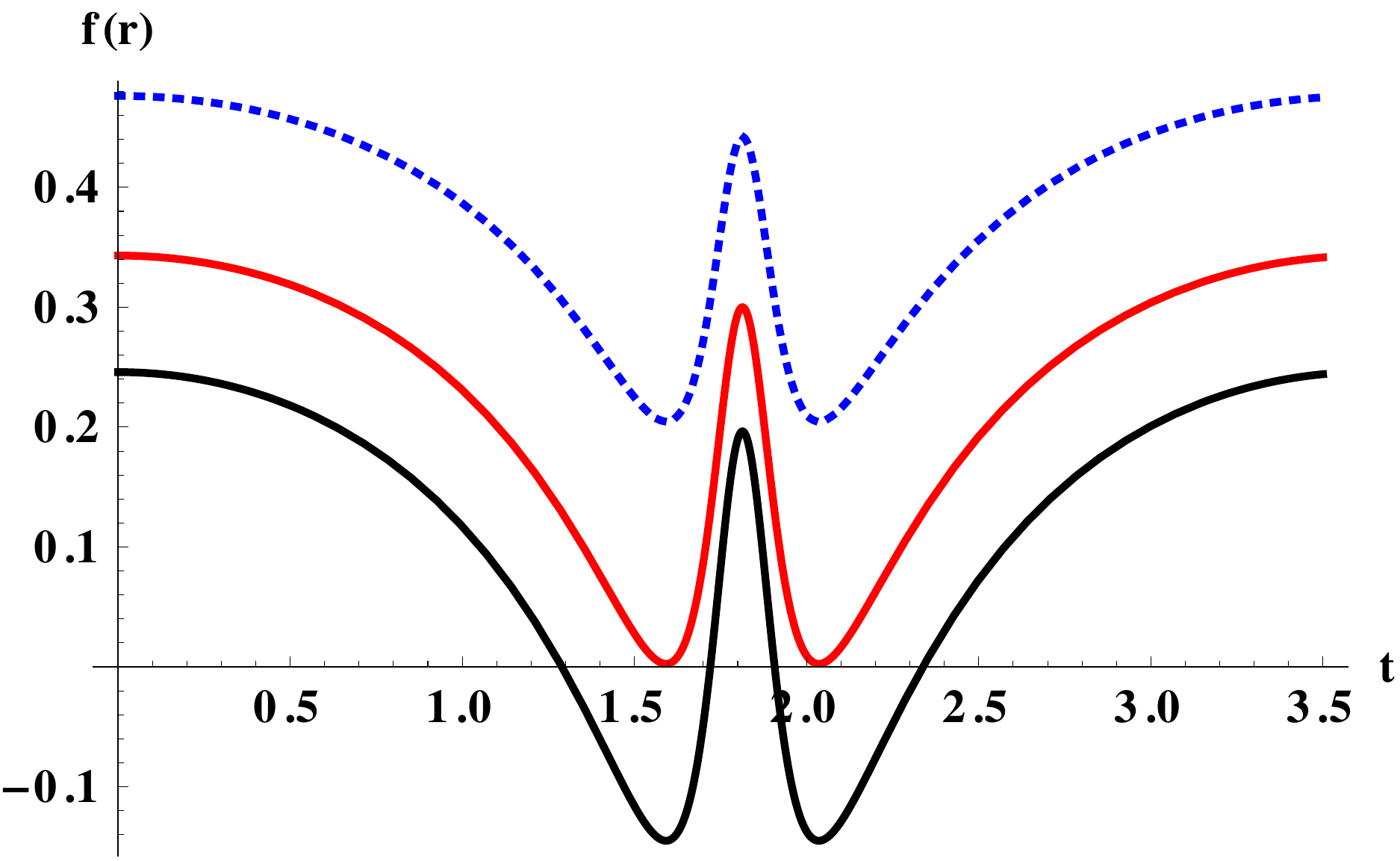}
\caption{Plot of $f(t)$ for different values of $r_{0}$ }
\label{internal}
\end{minipage}
\end{figure}

Finally, we analyse  the horizon structure of the exterior 
geometry as well as the issue of apparent horizons of the interior collapsing spacetime. The apparent horizon indicates 
the boundary of the  trapped surfaces, and is the location where the normal 
vector to the $Q=constant$ hypersurface (see eq.(\ref{gencoll})) becomes null. For the quantum corrected 
internal collapsing solution, this condition reduces to $\bar{g}_{\mu\nu}Q_{,\mu}Q_{,\nu}=0$. 
From eqs.(\ref{QFRW}) 
and (\ref{ata}), we then find that the apparent horizon curve in the $r-t$ plane is
\begin{equation}
r_{ah}(t)=\frac{cN(t)}{\dot{A}}=\frac{c^{2}\sqrt{A(t)}}{G}\bigg[\mathcal{E}-\frac{\Lambda c^{4}}{3G}A^{3}(t)
-\frac{c\hbar}{2}\sqrt{\frac{\Lambda}{3}}\Big(8 +18 \Big(\frac{\mathcal{C}_{1}}{A(t)}\Big)^{5}\Big)\bigg]^{-1/2}~.
\label{Aaph}
\end{equation} 
%By evaluating the above expression at $t=0$ one can see that, whether the apparent horizon exists at the 
%start of the collapse depends on the the values of $\mathcal{E}$, $\Lambda$ and $\mathcal{C}_{1}$. 
Analytical considerations of eq.(\ref{Aaph}) are indeed cumbersome, and we will resort to
a graphical analysis. As an illustration, as before, we 
choose $\mathcal{C}_{1} = 0.01$, with $G=c=\hbar = \Lambda = 1$. 
The apparent horizon curve of eq.(\ref{Aaph}) is plotted in fig.(\ref{fig:app horizons}), 
for three different values of the constant  $\mathcal{E}=15$ (solid red), $20$ (dotted blue) and $25$ (dashed black), 
where the blue and black curves are scaled by a factor of $1.5$, for ready comparison. 
As can be seen all these curves have a minimum at certain points of time, before and after the bounce. This 
indicates that there exists a minimum value of $r=r_{(ah)min}$ below which apparent horizon never forms \cite{BMM}. 
Thus if the matching radius of the internal metric $r_{0}< r_{(ah)min}$, the apparent horizon never forms in the 
collapsing cloud. For future, use we record the expression $r_{(ah)min}\simeq 0.034$ for
${\mathcal E} = 15$ (the dotted blue curve in fig.(\ref{fig:app horizons})). 

Now from the exterior solution of eq.(\ref{ext}), we glean that the only singularity there is at ${\mathbf r}=0$. 
This can be seen, for example, by considering the Ricci scalar which, from eq.(\ref{Ric}), behaves as ${\mathbf r}^{-8}$, and the 
Kretschmann scalar which goes as ${\mathbf r}^{-16}$ for small ${\mathbf r}$. The full solution with 
the internal metric of eq.(\ref{QFRW}) and the external one of eq.(\ref{ext}) will thus have no naked singularity
with a non-zero matching radius. 
The horizon structure of the exterior depends on the zeros of the function $ f_{q}(\mathbf{r})$, i.e., the 
roots of the algebraic equation  $\frac{\Lambda}
{3}\mathbf{r}^{8}+\mathbf{r}^{6}-2\frac{G\mathcal{M}}{c^{2}}\mathbf{r}^{5}+\mathcal{C}_{2}r_0^8L_{p}^{2}=0$. 
This has eight roots and we will be interested in a situation where the constants can 
be chosen in such a way that at most two of them are real and positive. As before, we
choose $\mathcal{C}_{1} = 0.01$ (with $G=c=\hbar = \Lambda=1$). Then, ${\mathcal C}_2$ is obtained from
the second relation of eq.(\ref{C1C2}), and ${\mathcal M}$ from eq.(\ref{MExp}), where we will set 
${\mathcal E} = 15$, as we did for fig.(\ref{denpr}). 

Now with these constants, setting $f_{q}(\textbf{r})=0$, we obtain the horizon structure of the exterior geometry
with at most two real roots. For different values of $r_0$, we will have either a) two real positive roots (inner and outer horizons), b) One
real positive root (the extremal solution) and c) no real positive root, which is a regular  bouncing compact object. 
These should correspond to three cases, i.e., $r_0 >, =, {\rm or} < r_{(ah)min}\simeq 0.034$. 
In fig.(\ref{fig:inner and outer}), we illustrate these three cases by plotting the function 
$f(\mathbf{r})$. In this figure, the red curve corresponds 
to $r_{0}=r_{(ah)min}$, the dotted blue curve for $r_{0}=0.03<r_{(ah)min}$, and the dashed black curve 
for $r_{0}=0.036 > r_{(ah)min}$. As can be seen, when $r_{0}>r_{(ah)min}$, there are two horizons in the exterior but when 
$r_{0}<r_{(ah)min}$, the exterior geometry has no horizon. Finally, $r_{0}=r_{(ah)min}$  indicates the extremal case.
%%%%%%%%%%%%%%%%%%%%%%%%
%%%%%%%%%%%%%%%%%%%%%%

Note that in the case of two horizons, since the $\mathbf{r}-\mathbf{r}$ 
component of the metric diverges at the inner and outer horizons, the coordinates $(\mathbf{t},\mathbf{r},\theta, \phi)$ 
are unable to describe the full spacetime geometry of the (static) solution. To describe the full geometry one has to  
introduce Kruskal like coordinates which will give the maximal extension of the spacetime.  
However in our case, this problem can be avoided by joining the spacetime smoothly with an interior 
solution at a radius greater than that of the outer horizon.  

Also, with two horizons (the case $r_{0}>r_{(ah)min}$), the work of \cite{ABMG} (see also \cite{JM}) suggests that 
the collapsing sphere crosses both the outer and the inner horizons, and 
finally bounces back, and this process is accompanied by the formation of a trapped surface and an apparent horizon. 
In that case, for a bouncing solution to exist, the formation of the inner horizon is crucial. As shown in \cite{ABMG}, 
there exists a 
minimum value of the radius which the surface of the collapsing matter can achieve. The sphere can bounce only 
in an untrapped region of the spacetime. In other words, the collapsing sphere can bounce only at a certain 
point of its evolution which is determined by the horizon structure of the exterior. 

Now, it can be shown sraightforwardly from the results of \cite{ABMG} that a necessary condition for the bounce 
described above is that the quantity $f(t) >0$ at the time of bounce, according to a co-moving observer
on the matching hypersurface. In fig.(\ref{internal}), we plot the function $f({\mathbf r}_0 = r_0A(t))$ for the
values of $r_0$ given in fig.(\ref{fig:inner and outer}) with the same color coding. It is seen that this condition is indeed satisfied
in our case. There is, in fact, an upper limit of $r_0$ beyond which this condition is not satisfied. 
In this case, we find that $r_0 \lesssim 0.04$. 

In the case  $r_{0}<r_{(ah)min}$, no horizons 
exist anywhere in the collapsing sphere as well as the exterior, and the bounce occurs in an untrapped region.
The extremal solution with $r_{0}=r_{(ah)min}$ has a single horizon and the bounce again occurs in 
an untrapped external region. For the last two cases also, the condition $f(t) >0$ is satisfied at the
time of bounce, as seen from fig.(\ref{internal}). 

\section{Conclusions and Discussions}
\label{Conc}

In this paper, we have explicitly demonstrated the role of the quantum potential in a semi-classical
treatment of the Wheeler-De Witt equation,  
using the functional Schrodinger formalism for realistic collapse models. Using this formalism for solving the
Schrodinger equation with a position dependent mass, we computed the
wave functions for marginally bound dust collapse, in the absence of radiation. 
Next, we performed a de Broglie-Bohm analysis of the functional Schrodinger 
equation (which, from the point of view of the comoving observer is a position dependent mass equation), and 
using a nonstationary state wave function, explicitly calculated the quantum potential. The quantum potential acts 
as a source of perfect fluid energy momentum tensor so that the particle trajectories obtained from integrating 
the velocity equation (near the classical singularity) never reach the classical singularity. 
We computed a quantum corrected metric which is related to the original collapsing metric via a purely disformal
transformation, and modifies the original Hamiltonian by the addition of the quantum potential. This analysis
yields the explicit forms of the energy density and the pressure, and we could glean novel insight into the
role of the quantum potential in the collapse scenario.
The collapsing dust shell bounces 
back after a minimum radius where the effect of the quantum potential is maximum. The motion of the shell shows 
an oscillatory behavior instead of falling into the classical singularity. Further, we computed a quantum corrected
exterior metric and established that the bounce satisfies required physical conditions.

To connect with important existing literature, and to contrast our methods with these, 
we mention that in \cite{SD}, Bohmian mechanics was used to add   
quantum correction terms in the Raychaudhuri equation and it was argued, using the well known
property of Bohmian trajectories, that they do not
cross, i.e., unlike the usual particle  geodesics in GR they do not
form any caustic. Thus even if the spacetime has singularities, the
particle trajectories will never reach them. This quantum Raychaudhuri equation
was later used in \cite{AD} to resolve the cosmological singularity in
the FRW model. Further, in \cite{BMM}, in the framework of classical homogeneous
dust and the radiation FRW model, it was shown by using an effective profile of the energy
momentum tensor (whose form is inspired by loop quantum cosmology
models) that one can avoid the classical singularity. The quantum
effects were shown there to produce a negative pressure, which causes the shell to bounce
back instead of falling into the singularity.  Similar bouncing behavior
was also noticed by others in the quantum version of  inhomogeneous  dust
models (see \cite{RB}, \cite{LMMB}). Our method on the other hand is novel, and distinct 
from these in that it provides an explicit evaluation of the wave function, and a quantum 
corrected metric.
As an immediate future application, it would be interesting to analyse a similar situation with a positive cosmological
constant.

\begin{center}
{\bf Acknowledgements}
\end{center}

\noindent
We sincerely thank Saurya Das for comments on a draft version of this paper. We acknowledge the anonymous referee
for valuable comments. The work of SC is supported by CSIR, India, 
via grant number 09/092(0930)/2015-EMR-I.

\end{document}